# A bibliometric analysis of Canadian LIS scholars and practitioners' research contributions

Jean-Sébastien Sauvé[1]*, Madelaine Hare[2], Geoff Krause[3], Constance Poitras[4], Poppy Riddle[5], Philippe Mongeon[6]

[1]jean-sebastien.sauve@umontreal.ca
https://orcid.org/0000-0002-9472-2965
École de bibliothéconomie et des sciences de l'information, Université de Montréal, Canada

[2]maddie.hare@uottawa.ca
https://orcid.org/0000-0002-2123-9518
Digital Transformation and Innovation, University of Ottawa, Canada

[3]gkrause@dal.ca
https://orcid.org/0000-0001-7943-5119
Department of Information Science, Dalhousie University, Canada

[4]constance.poitras@umontreal.ca
https://orcid.org/0000-0002-3545-696X
École de bibliothéconomie et des sciences de l'information, Université de Montréal, Canada

[5]pnriddle@dal.ca
https://orcid.org/0000-0001-7862-7848
Department of Information Science, Dalhousie University, Canada

[6]PMongeon@dal.ca
https://orcid.org/0000-0003-1021-059X
Department of Information Science, Dalhousie University, Canada
Centre interuniversitaire de recherche sur la science et la technologie (CIRST), Université du Québec à Montréal, Canada

* Corresponding author (jean-sebastien.sauve@umontreal.ca)

## Abstract

Canada's research productivity in Library and Information Science (LIS) is significant: studies have found that Canada ranks third globally in terms of output. As the LIS field continues to grow, the pace of output accelerates, and the scope of this work expands. The recently launched *Canadian Publications in Library and Information Science Database* compiles all Canadian scientific publications, including those authored by faculty members and academic librarians. This database offers the advantage of encompassing articles and librarian publications that may not be typically included in traditional bibliometric surveys, such as those conducted using databases like Web of Science, Scopus, and Library and Information Science Abstracts (LISA). Using this data, this study maps the scholarly contributions of Canadian LIS scholars and academic librarians to the field of LIS and examines whether Canadian LIS research is characterized by silos. This paper examines the similarities and differences in research output, impact, topics, and publication venues between academic librarians and scholars in Canada, as well as the extent to which academics and practitioners engage in research collaborations or reference each other's work. We find that while there is some degree of overlap in research topics and publication venues between LIS academics and academic librarians, the two groups appear to act as distinct research communities with distinct topical foci and publishing habits. The two groups also do not appear to engage with each other strongly, either through collaboration or citing each other's work.

# 1. Introduction

Canada's research productivity in Library and Information Science (LIS) is significant, with the country contributing 4.48% of worldwide publications and ranking third in global output (Jabeen et al., 2015). Engagement in research activities is considered a key aspect of the workload for LIS faculty members at Canadian institutions. As the roles of academic librarians have expanded over time, this has led to an increasing emphasis on research as part of their broader responsibilities (Fox, 1997a; Ducas et al., 2020). Fox (2007c) revealed that more than half of Canadian academic librarians are either required or encouraged to engage in scholarly activities as part of their job. Job titles such as "research librarian", "scholarly communication librarian", and "digital publishing/open scholarship librarian" have emerged and are frequently mentioned in academic librarians' job descriptions (Kandiuk & Sonne de Torrens, 2018), highlighting their growing recognition as important contributors to institutional research output.

While LIS academics and librarian practitioners share topical interests (White & Cossham, 2017), research produced by each group sometimes evolves in parallel. This has resulted in an observable topical distinction between practitioners, generally focused on pragmatic issues (Hall & McBain, 2014), and theorists, who are less involved in field practice (Clayton, 1992; Genoni et al., 2006). Nguyen and Hider (2018) found such disparities, noting that librarians tend to focus their research on issues more relevant to practical knowledge. Methodologies have also been found to vary, with researchers in academic contexts utilizing a more diversified range of methods than those in professional settings (Hildreth & Aytac, 2007). Ardanuy & Urbano (2017) observed that LIS is an inherently practice-based field and recognized that knowledge produced about professional practice (particularly its transformation) should be in high demand.

Evidence of a lack of collaboration suggests that silos in LIS research exist. Global bibliometric studies in information science underline the low number of research collaborations between faculty members and librarians. Hildreth and Aytac (2007) observed that topical gaps between librarians and academics were minimal, but so was their tendency to collaborate. Finlay et al. (2013) found that less than 3% of their corpus (n=4,772) resulted from faculty and librarians' collaboration. Babb (2017) surveyed and interviewed librarians at Canadian universities and found that while they mostly conducted LIS-oriented research, they also desired to topically diversify from the subject area for which they liaised and worked collaboratively with academic researchers (Babb, 2017). Moreover, Borrego et al. (2018) observed that instances of library-affiliated authorship in non-LIS journals were increasing and found evidence of collaboration between librarians and faculty, though noted that further investigation into their role in the research process and inter-institutional collaborative behaviours is needed.

Though faculty members and librarians traditionally engage at the interface of scholarly production, greater collaboration between faculty members and librarians is highly desirable for the advancement and vitality of the discipline. As stated by Ponti (2013), collaboration "can create stronger academic-practice networks. Even after a project is completed, interpersonal relationships remain and constitute a social capital from which to draw to initiate new projects" (p. 34). Few studies have addressed the scope of LIS research in Canada and the collaboration behaviours of LIS academics and practitioners. Paul-Hus and Mongeon (2016) examined the scientific production of faculty members from the eight LIS schools in Canada. Mongeon et al. (2023) further mapped the research conducted by faculty members, doctoral students, and postdoctoral fellows from various Canadian LIS schools. However, both studies were unable to include academic librarians in the analysis.

## 2. Research objectives

This study aims to address existing silos to facilitate greater use of LIS literature produced by both Canadian academic librarians and faculty members. Its primary objectives are 1) to map the scholarly contributions of Canadian faculty and academic librarians in the field of LIS and 2) subsequently examine whether Canadian LIS research is characterized by silos. We use a comprehensive database of all Canadian LIS publications authored by faculty members and academic librarians created as part of the SSHRC-funded project "Breaking down research silos" (Sauvé et al., 2024a). This database offers the advantage of encompassing articles not typically included in traditional bibliometric surveys, such as those conducted using databases like Web of Science (WoS), Scopus, and Library and Information Science Abstracts (LISA).

Specifically, this paper aims to answer the following research questions:

**RQ1** What are the similarities and differences in publications, publication venues, and topics between Canadian LIS faculty members and academic librarians?

**RQ2** To what extent do Canadian LIS faculty members and academic librarians engage in research collaborations or reference the other group's work?

By addressing these questions, this investigation aims to bridge gaps and foster collaboration between academic librarians and scholars in the Canadian LIS research landscape.

This paper first reviews the literature on the research tasks undertaken by Canadian academic librarians and their scientific production and contributions to LIS, which have often been overlooked. It will examine the types of research conducted, publication trends, barriers and incentives faced in research endeavours, and the extent of collaboration between librarians and scholars. This analysis will provide the foundation for presenting the bibliometric tools employed to address our research questions and the corresponding results. The discussion of our findings compares the output of Canadian academic librarians with that of LIS faculty members, evaluating whether each group's scientific production occurs in relative isolation. The telos of this study is to shed light on the potential existence of silos and their implications for the research landscape in Canada.

## 3. Literature review

### 3.1 Academic research in LIS

Historically, LIS research became more institutionalized in the 1970s as LIS academic courses and programs proliferated (Saracevic, 1979). At the same time, researchers from different disciplines began to take an interest in the study of information, influencing the development of the discipline and leading to LIS being widely regarded as an interdisciplinary or multidisciplinary discipline (Aharony, 2011; Onyancha, 2018; Zhang et al., 2023), but also as a fragmented discipline (Vakkari, 2024). This influence has continued, while several recent studies confirm the growing contribution of other disciplines to scholarly production in LIS (Chang, 2018, 2019; Lund, 2020; Ma & Lund, 2021; Vakkari et al., 2022, 2023). Moreover, numerous empirical studies have noted a gradual shift in the topical landscape of the field, where topics related to librarianship have declined in favour of topics such as information retrieval, knowledge management, and bibliometrics (Figuerola et al., 2017; Larivière et al., 2012; Vakkari, 2024). At the same time, the methods used by LIS researchers have diversified as data availability has increased, with data mining, bibliometrics, experiments, and surveys as dominant strategies (Matusiak et al., 2024). While LIS is viewed as an applied discipline concerned with practical problems, therefore benefitting from strong links with practitioners (Petras, 2023), a distance between researchers and practitioners in information science began to develop in the late 1970s (Saracevic, 1979).

Because LIS faculty members are established as researchers due to their workload expectations and the historical positioning of researchers within departments and schools in the academy, this literature review turns focus to librarians as researchers. It overviews their current status and duties in relation to research at Canadian higher education institutions, publication types they most commonly contribute to producing, and barriers and motivations to their involvement in research activities. With this positioning, we then describe what is currently known about research collaborations between faculty and librarians to contextualize the findings of this investigation.

### 3.2 Librarians' status and research duties in Canada

Eight of Canada's 95 publicly funded universities are home to a LIS school: Dalhousie University, McGill University, Université de Montréal, University of Alberta, University of British Columbia,

University of Ottawa, University of Toronto, and University of Western Ontario (Universities Canada, n. d.). While the duties of faculty members from these schools are usually well established when it comes to research and teaching, this distinction does not apply to academic librarians, as they can possess different statuses.

Zavala Mora et al. (2023) highlight the Downs Report of 1967, which outlined recommendations for modernizing Canadian research and university libraries. This report recognized the importance of granting academic status to librarians and acknowledged the value of their engagement in research activities. By proposing measures such as freeing librarians from certain professional responsibilities, the report aimed to support the development and advancement of the profession. Since these recommendations have not been fully implemented, similar assertations have been made recently by professional associations. The Canadian Association of Research Libraries (CARL) (DeLong et al., 2020), the Canadian Association of University Teachers (CAUT) (CAUT, n. d.), and the Canadian Association of Professional Academic Librarians (CAPAL) (CAPAL, 2020) recognize the significance of librarians' research responsibilities, considering them as relevant and important as those of their faculty colleagues (Schrader et al., 2012; Zavala Mora et al., 2023). Further, many librarians conduct research in collaboration with colleagues within the same professional associations.

In contrast to academic roles, where the allocation of time for research is relatively standardized across institutions, there is "no common consensus on the norms for librarian's time commitment to scholarship" (Fox, 2007b, p.15). According to Fox's survey of Canadian librarians (2007c), research was considered optional, unnecessary, or even discouraged for about half of the respondents. Similarly, Berg et al.'s (2013) survey of librarians and administrators at CARL member institutions revealed inconsistencies in the expectations of academic librarians regarding scientific production: while some institutions required research for tenure advancement, others did not specify the nature of such research activity or whether it was expected at all (Berg et al., 2013). The study also highlighted a discrepancy between perceived expectations for research production between administrators and librarians, with the former often rating expectations as low or adequate, while the latter considered them to be adequate or too high (Berg et al., 2013). Kandiuk and Sonne de Torrens (2018) reviewed the terms of employment in university collective agreements and policies and conducted a survey of Canadian academic librarians. Their findings indicated that most universities (72%) made provisions for librarians' academic freedom to conduct research and produce scholarship through explicit statements. Over half (52%) required research, although definitions of research and scholarship varied. Additionally, 19% of collective agreements and policies specifically stipulated that research must be conducted in the LIS field and professional practice. Despite varying expectations of their research activities, Canadian librarians demonstrate a strong interest in conducting and engaging in research, which is closely tied to personal development within the library and archive professions (Doucette & Hoffmann, 2019).

Fox (2007c) draws attention to the unique position of Québec within Canada. Academic librarians in Québec typically do not have the same designation as faculty members, and a subsequent absence of certain research benefits, such as research sabbaticals. This distinction is evident in the situation of university librarians in French-speaking Québec universities as compared to their English-speaking counterparts (Zavala Mora et al., 2023). In French-speaking Québec, librarians hold professional status, which is distinct from academic status. For example, librarians at Université de Montréal are part of the support staff union, while at Université Laval, they belong to the professional union. These differences have significant implications for scientific productivity in LIS in Canada, particularly considering that there is only one French-speaking LIS school (Université de Montréal) and one bilingual LIS school (University of Ottawa) in Canada.

## 3.3 Publication types

According to the survey conducted by Fox (2007c), the research activities of librarians in Canada differ in focus from those of faculty members. Most research university librarians in Canada engage in activities such as presenting at conferences, writing unpublished reports, publishing in non-peer-reviewed journals, and contributing to blogs or professional websites. Peer-reviewed journal publications ranked sixth in terms of frequency, with approximately 45% of librarians having made at least one contribution in that category (Fox, 2007c). Sugimoto et al. (2014) found that librarians predominantly publish their research in conference papers, posters, and presentations, accounting for 65.9% of their overall scholarly output. While 54.2% of librarian respondents utilized peer-reviewed journals, these publications are less popular amongst practitioners due to concerns about the slow evaluation process and a preference for blogs to expedite the dissemination of findings (Sugimoto et al., 2014). Apart from time, 70% of respondents indicated that the target audience and potential impact of their work were significant considerations in their choice of journal (Sugimoto et al., 2014). More recently, a survey conducted by Hoffmann et al. (2023) indicates that conference presentations are by far the most popular form of publication, followed by peer-reviewed articles and posters.

The methods practitioners tend to use also vary from those used by faculty members. Slutsky and Aytac (2014) examined academic librarian output between 2008 and 2012 in four science, technology and medicine journals, finding that 84% of papers from their corpus used quantitative methods, while 10% used exclusively qualitative methods. Librarians published predominantly surveys and content analyses, as well as descriptive articles and evaluative and exploratory articles (Slutsky & Aytac, 2014). O'Brien and Cronin (2016) confirm the importance of case studies for librarian publications in their analysis of 93 peer-reviewed journal articles produced by library staff working in higher education colleges in Ireland, where one in every three papers used this approach. The same conclusion appears in Akers and Amos (2017) who argued that the case study, despite its low position in the hierarchy of evidence, is both popular among librarians and relevant since they "can communicate timely and innovative approaches to librarianship [and that] publishing a case study is an alternative way for practising librarians to engage in scholarly discourse" (p. 115).

## 3.4 Barriers and motivations to academic librarians' involvement in research

Literature on barriers to librarian participation in research is extensive. In a survey of 476 Canadian librarians at research institutions, Fox (2007b) reported that librarians spend only 7-8% of their time on research, which largely constituted their own time rather than comprising their workday. Berg et al. (2013) also found time to be a significant barrier for librarians, as well as a "lack of funds" and a "lack of support" from their institution. Similarly, Kennedy and Brancolini's (2012) identified factors which included time constraints, a lack of institutional, financial, and emotional support, self-confidence, familiarity with the research process and methods, research jargon, and motivation. Kennedy and Brancolini (2012) highlight a lack of research training in librarian education, though the LIS curriculum adequately prepares students to read and understand research articles. This was confirmed in a survey by Kennedy and Brancolini (2018), where only 17% of librarians said that their master's studies prepared them well for research. The issue of research training was also raised by Luo (2011), who noted, however, that shortcomings in methodology training had no impact on librarians' scientific output. The results of a questionnaire submitted to librarians at the University of Saskatchewan indicated that librarians are highly motivated to undertake research projects but that difficulties arise in defining research programs or using research methods (Schrader et al., 2012). Further, librarians face challenges in being recognized or negotiating credit by way of authorship for their research contributions to faculty projects (Bloss et al., 2022; Babb, 2017).

Canada-wide initiatives aimed at assisting librarians in meeting the research expectations and requirements of their institutions have been found lacking by Jacobs and Berg (2013) though initiatives aimed at supporting librarians' research activities have steadily grown over the past decade. A case study conducted at the University of Saskatchewan demonstrated program development's positive impact on enhancing librarians' knowledge and skills in research and scholarly communications (Shrader et al., 2012). McMaster University Library's "Faculty-Member-in-Residence" program involves non-librarian faculty members assisting librarians with their research (Detlor & Lewis, 2015). This program addressed challenges concerning librarians' research capabilities, such as lack of familiarity with the research process, limited support, and confidence and motivation, helping to cultivate a research-friendly environment. CARL has sponsored the Librarians' Research Institute (LRI), whose primary objective is to cultivate a robust intellectual culture among librarian researchers in Canada through a peer mentorship program and a diverse curriculum which includes balancing research and practice, understanding research processes and planning, adopting appropriate approaches and methodologies, and effectively disseminating research findings and making professional contributions (Jacobs and Berg, 2013). As of 2023, the LRI is in its seventh session, indicating an ongoing demand and necessity to support research education infrastructure for librarians in Canada (CARL, n.d.).

### 3.5 Collaboration between faculty and librarians

The benefits of collaboration between researchers and librarians are well documented and represent a growing trend. Analyses of research output by academic librarians in the United States have revealed that most publications originate from a small but growing number of authors and institutions. Despite an increase in the number of authors, there is a slight decline in the production of LIS-related contributions. This increase in authors suggests a rise in collaborative efforts rather than individual articles (Weller et al., 1999; Wiberley et al., 2017; Blecic et al., 2017). Canadian librarians' research output was examined by Julien and Fena (2018) through a content analysis of the *Canadian Journal of Library and Information Science* (CJILS/RCSIB), the oldest bilingual LIS journal in Canada, spanning a period of 31 years. The study found an upward trend in multiple authorships over time, though authors were not specified as librarians and faculty (Julien & Fena, 2018). There was also a noticeable topical shift, with a decreasing proportion of studies relating to information retrieval as compared to information behaviour. Moreover, empirical studies have remained dominant, with an increasing number of such studies in recent years (Julien & Fena, 2018).

Despite these trends, collaboration remains limited. Finlay and colleagues (2013) analyzed a corpus of approximately twenty journals spanning from 1956 to 2011 and found that librarians accounted for only 31% of all scientific communications, with a mere 3% resulting from collaboration between librarians and scholars. White and Cossham (2017) identified keywords to determine the subjects on which librarians and LIS scholars collaborated most frequently. They observed that only 6% of their corpus involved collaboration, while 40% of articles were authored solely by librarians and only 39% by faculty members. Slutsky and Aytac (2014) examined 574 articles from four journals focused on science, technology, and medicine between 2008 and 2012, finding that 166 were written exclusively by librarians, around 60 were the result of collaboration between librarians and researchers, and 71 were authored by researchers alone. These findings parallel the conclusions of Schlögl and Stock's (2008) seminal paper, which investigated the relationship between librarians and scholars in LIS scholarly communications through a citation analysis of ten German language and 40 international journals, as well as questionnaires sent to editors and readers. The authors concluded that professionals in the field are highly productive in scholarly communication, noting, "There might not be many disciplines where practitioners contribute as much to the knowledge base" (p. 650). However, their study also confirmed the presence of silos, as academics and librarians tended to remain within their respective communities with professional journals citing research-oriented journals infrequently, and vice versa (Schlögl & Stock, 2008). Consequently, the authors concluded that practitioners both publish in and read LIS scholarly journals, but the journals preferred by professionals differ from those in which researchers publish (based on factors such as editorial board size, number of references per article, bibliography, half-life of references, and presence of advertising), and limited exchange occurs between professionals and academics (Schlögl & Stock, 2008).

## 4. Data and methods

We used the open dataset of Canadian LIS publications by Sauvé et al. (2024a), which contains publications authored by researchers affiliated with one of the eight ALA-accredited degree-granting academic units in Canada (academics) and by librarians working in Canadian universities (Sauvé et al., 2024b). We limited our analysis to articles, notes, reviews, books, book chapters, conference papers, letters, book reviews, and editorial materials indexed in OpenAlex, for a total of 9,261 publications out of the 13,775 total publications included in the Sauvé et al. (2024a) dataset.

We assigned each paper to one of three groups (academic, collaboration, practitioner) based on the status of the authors indicated in the Sauvé et al. (2024a) dataset. A paper is assigned to the academic group if at least one author is a Canadian LIS academic and none of the authors are Canadian LIS practitioners. A paper is assigned to the practitioner group if at least one Canadian LIS practitioner and no Canadian LIS academics are listed as authors. A paper is assigned to the collaboration category if it has at least one Canadian LIS academic and one Canadian LIS practitioner listed on the byline. Our dataset includes 6,178, 2,935 and 148 publications in the academic, practitioner, and collaboration groups, respectively.

We were interested in exploring similarities and differences in research topics across groups, so we used the topic classification of OpenAlex, retrieved from their API with the openalexR package (Aria et al., 2024). It is based on the work of a hierarchical classification that uses citation-based clusters to assign individual papers to domains, fields, subfields, and topics (Van Eck, 2024). In OpenAlex, a paper can be assigned to multiple topics, each with a score representing the strength of the association between the paper and the topic. Because some of the topics are similar and tend to co-occur, we only kept the topic with the highest score for our analysis (so each paper has a single topic). The resulting dataset, used to answer RQ1, contains the following information for each observation:

- Document type
- Venue
- Group (academic, practitioner, collaboration)
- Topic

Our second research question investigates interactions between the two groups (academics and practitioners) using co-authorship and citations. The co-authorship analysis was done using the *authors_publications* and *authors* tables from the dataset; the former links authors with publications and was thus used to create a network file where nodes represent individual authors and edges represent co-authorship weighted by the number of co-authorships in the dataset. Each node is assigned to a group (academic or practitioner) based on the classification of Sauvé et al. (2024a). The resulting collaboration network was visualized with Gephi (Bastian et al., 2009), an open-source network analysis and visualization software.

The dataset includes a table of citation links between the papers in the dataset obtained from OpenAlex, which was used to create a matrix with the number and percentage of citations within and between each publication group (academic, practitioners, collaborations).

We used custom R scripts (R Core Team, 2023) to perform all data processing and analysis.

## 5. Results

### 5.1 Outputs

Table 1 reveals that of the 9,261 publications in our dataset, the majority are journal articles (71.3%), followed by reviews and book chapters. Academics tended to publish at higher rates, authoring 6,178 (66.7%) of total publications, while practitioners authored 2,935 (31.7%).

Collaborations accounted for only 137 publications, or less than 2% of total output. Articles and conference papers were the most common formats in which collaboration manifested. One highlight of the results presented in Table 1 is the substantially higher proportion of reviews and book reviews performed by practitioners (23.3% and 6.4%, respectively) compared to academics (11% and 2.1%, respectively).

Table 1. Number of publications by group and document type.

| Document type | Total N | Total % | Academic N | Academic % | Practitioner N | Practitioner % | Collaboration N | Collaboration % |
|---|---|---|---|---|---|---|---|---|
| Article | 6,601 | 71.3 | 4,600 | 74.5 | 1,898 | 64.7 | 103 | 69.6 |
| Review | 1,022 | 11.0 | 323 | 5.2 | 685 | 23.3 | 14 | 9.5 |
| Book chapter | 667 | 7.2 | 586 | 9.5 | 73 | 2.5 | 8 | 5.4 |
| Conference paper | 497 | 5.4 | 431 | 7.0 | 45 | 1.5 | 21 | 14.2 |
| Book review | 191 | 2.1 | 4 | 0.1 | 187 | 6.4 | 0 | 0.0 |
| Editorial material | 105 | 1.1 | 89 | 1.4 | 15 | 0.5 | 1 | 0.7 |
| Note | 78 | 0.8 | 57 | 0.9 | 20 | 0.7 | 1 | 0.7 |
| Book | 66 | 0.7 | 60 | 1.0 | 6 | 0.2 | 0 | 0.0 |
| Letter | 34 | 0.4 | 28 | 0.5 | 6 | 0.2 | 0 | 0.0 |
| Total | 9,261 | 100 | 6,178 | 100 | 2,935 | 100 | 148 | 100 |

## 5.2 Venues

The top venues of publications by academics (Table 2) show that the conference proceedings of two prominent LIS research organizations supporting both researchers and practitioners (*CAIS* and *ASIST*) are among the top five venues. The list includes several LIS-specific journals (e.g., *Library & Information Science Research*, *Cataloging & Classification Quarterly*, *Information Communication & Society*, etc.), but also multidisciplinary venues such as *PLoS ONE* and *First Monday*, and venues from disciplines other than LIS, some with obvious ties to the field (e.g., *Lecture notes in computer science*) and others for which this link is less obvious (e.g., *Marine Pollution Bulletin*).

Table 2. Top 20 venues of publications by academics

| Venue | N. pubs | Pct. pubs |
|---|---|---|
| Proceedings of the Association for Information Science and Technology | 289 | 4.7 |
| Lecture notes in computer science | 177 | 2.9 |
| Proceedings of the Annual Conference of CAIS | 168 | 2.7 |
| Journal of the Association for Information Science and Technology | 156 | 2.5 |
| Routledge eBooks | 92 | 1.5 |
| Documentation et bibliothèques | 74 | 1.2 |
| Scientometrics | 69 | 1.1 |
| PLoS ONE | 66 | 1.1 |
| Library & Information Science Research | 65 | 1.1 |
| IGI Global eBooks | 62 | 1.0 |
| Journal of Documentation | 58 | 0.9 |
| Education for Information | 54 | 0.9 |
| Springer eBooks | 52 | 0.8 |
| Cataloging & Classification Quarterly | 48 | 0.8 |
| Elsevier eBooks | 43 | 0.7 |
| Information Communication & Society | 41 | 0.7 |
| Information Processing & Management | 38 | 0.6 |
| First Monday | 33 | 0.5 |
| Knowledge Organization | 32 | 0.5 |
| Lecture notes in business information processing | 31 | 0.5 |

The top venues practitioners in our dataset publish (Table 3) in are almost exclusively different from those of academics, save the multidisciplinary venue *PLoS ONE*. There is a high representation of venues focused on reviews, such as *Systematic Reviews*, *Cochrane library*, and *JBI Evidence Synthesis*. Venues are also more topically focused on librarianship as a professional practice, yet range across disciplines, with several dedicated to health librarianship (e.g. *Journal of the Canadian Health Libraries Association*, *Journal of the Medical Library Association*), including review journals, a methodology common in the health sciences. A diverse range of fields are also represented, such as youth and children's librarianship (e.g. *The Deakin Review of Children's Literature*), music (e.g. *CAML review*), and economics (e.g. *The American Economist*).

Table 3. Top 20 venues of publications by practitioners

| Venue | N. pubs | Pct. pubs |
|---|---|---|
| Evidence Based Library and Information Practice | 145 | 4.9 |
| The Deakin Review of Children s Literature | 130 | 4.4 |
| Partnership: The Canadian Journal of Library and Information Practice and Research | 68 | 2.3 |
| BMJ Open | 63 | 2.1 |
| Journal of the Canadian Health Libraries Association | 61 | 2.1 |
| The Journal of Academic Librarianship | 54 | 1.8 |
| College & Research Libraries | 43 | 1.5 |
| Systematic Reviews | 41 | 1.4 |
| Journal of the Medical Library Association | 38 | 1.3 |
| PLoS ONE | 33 | 1.1 |
| Cochrane library | 27 | 0.9 |
| Reference Services Review | 27 | 0.9 |
| CAML Review | 20 | 0.7 |
| JBI Evidence Synthesis | 20 | 0.7 |
| The Serials Librarian | 20 | 0.7 |
| College & Research Libraries News | 19 | 0.6 |
| Library Hi Tech | 19 | 0.6 |
| College & Undergraduate Libraries | 16 | 0.5 |
| The American Economist | 16 | 0.5 |
| Proceedings of the Annual Conference of CAIS | 15 | 0.5 |

Collaborative publications' venues (Table 4) overlap with venues preferred by academics or practitioners separately. However, more are common to practitioners (7) than academics (6) though this is relatively equal, with the addition of new venues not represented in the top venues of either group.

Table 4. Top 20 venues of publications co-authored by academics and practitioners.

| Venue | N. pubs | Pct. pubs |
|---|---|---|
| Proceedings of the Annual Conference of CAIS | 25 | 16.9 |
| First Monday | 6 | 4.1 |
| Proceedings of the Association for Information Science and Technology | 5 | 3.4 |
| The Journal of Academic Librarianship | 5 | 3.4 |
| Archives | 4 | 2.7 |
| BMJ Open | 4 | 2.7 |
| Journal of Documentation | 3 | 2.0 |
| Journal of the Association for Information Science and Technology | 3 | 2.0 |
| Journal of the Medical Library Association | 3 | 2.0 |
| Lecture notes in computer science | 3 | 2.0 |
| Library & Information Science Research | 3 | 2.0 |
| Partnership: The Canadian Journal of Library and Information Practice and Research | 3 | 2.0 |
| Archivaria | 2 | 1.4 |
| BMC Medical Research Methodology | 2 | 1.4 |
| BMC Medicine | 2 | 1.4 |
| Cataloging & Classification Quarterly | 2 | 1.4 |
| College & Research Libraries | 2 | 1.4 |
| College & Research Libraries News | 2 | 1.4 |
| Études de communication/Études de communication | 2 | 1.4 |
| Implementation Science | 2 | 1.4 |
| Journal of Information Policy | 2 | 1.4 |
| Journal of the Canadian Health Libraries Association | 2 | 1.4 |
| Knowledge Organization | 2 | 1.4 |
| Online Information Review | 2 | 1.4 |
| Presses de l'Université du Québec eBooks | 2 | 1.4 |
| Serials Review | 2 | 1.4 |
| The International Journal of Information Diversity & Inclusion (IJIDI) | 2 | 1.4 |

Figure 1 depicts the distribution of venues by share of articles authored by academics. The blue line is the theoretical distribution in which every venue has a proportional representation of academics and practitioners. It is apparent that approximately 37.5% of venues exclusively publish academics, and about 10% of venues exclusively publish practitioners (as indicated by the 0% of academics). This shows that about half of the journals in our dataset publish exclusively works by one of the two groups.

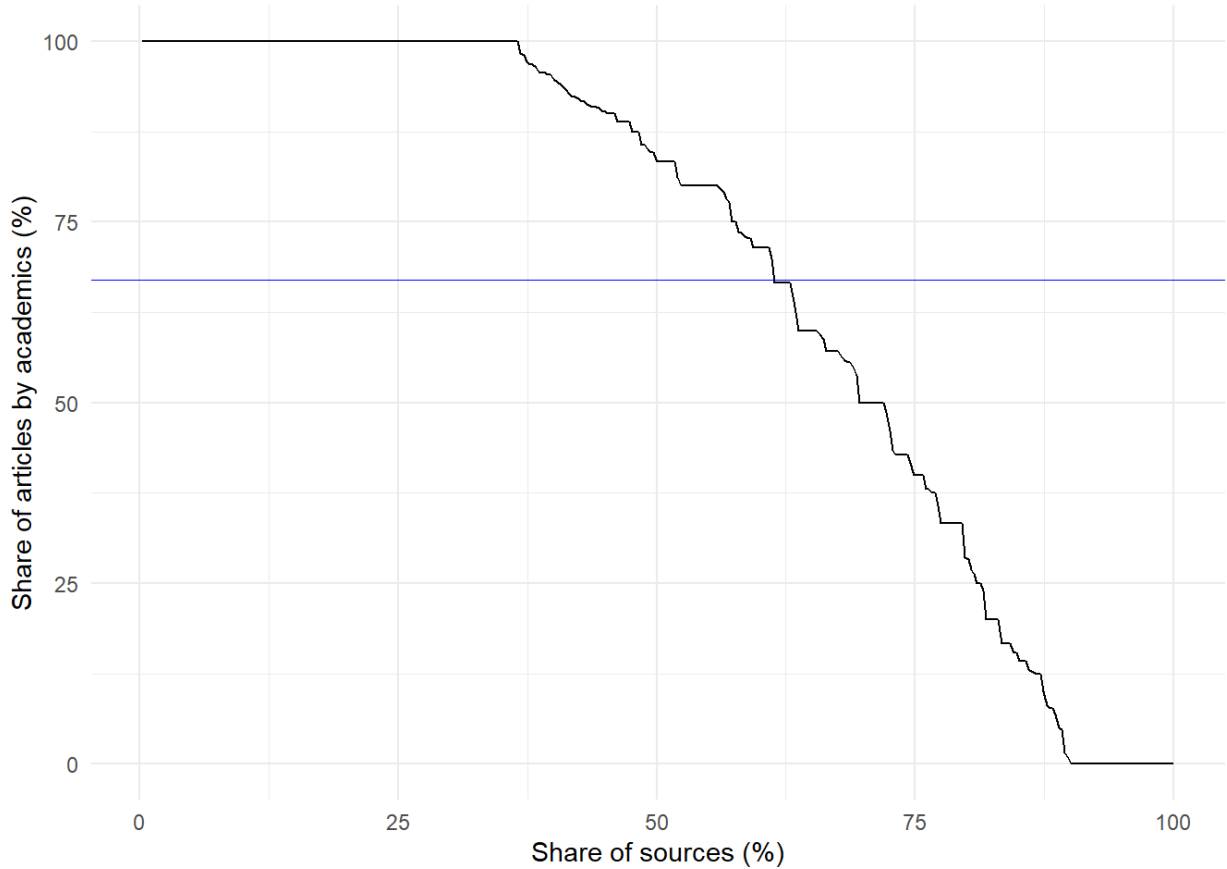

Figure 1. Distribution of venues by share of articles authored by academics.

## 5.3 Topics

The top topics of academic-authored publications are shown in Table 5. Topics relating to Scientometrics and Bibliometrics research and analysis amount to 3.6% of publications. Several topics relating to digital media, communications, and virtual knowledge sharing (i.e., *The Impact of Digital Media on Public Discourse*, *Digital Communication and Information Studies*, *Knowledge Sharing in Virtual Communities*, and *Digital Games and Media*) appear among the top topics, making up around 3.3% of publications. We also observe the prominence of core library science topics such as information literacy, information retrieval, and information behaviour. Archives are also represented in the *Archival Science and Digital Preservation* (58 publications) and *Digital and Traditional Archives Management* (41 publications) topics. The semantic web is also a notable inclusion in the list.

Table 5. Top 20 topics of publications authored by academics

| OpenAlex topic | N. pubs | Pct. Pubs |
| --- | --- | --- |
| Scientometrics and bibliometrics research | 123 | 2.0 |
| Bibliometric Analysis and Research Evaluation | 98 | 1.6 |
| Social Inclusion in Library Services for Newcomers | 85 | 1.4 |
| Knowledge Management and Sharing | 65 | 1.1 |
| Information Literacy in Higher Education | 58 | 1.0 |
| The Impact of Digital Media on Public Discourse | 61 | 1.0 |
| Digital Communication and Information Studies | 60 | 1.0 |
| Archival Science and Digital Preservation | 58 | 1.0 |
| Semantic Web and Ontologies | 55 | 0.9 |
| Social Media and Politics | 53 | 0.9 |
| Library Science and Information Literacy | 47 | 0.8 |
| Library Science and Administration | 50 | 0.8 |
| Semantic Web and Ontology Development | 50 | 0.8 |
| Knowledge Sharing in Virtual Communities | 44 | 0.7 |
| Information Retrieval Techniques and Evaluation | 43 | 0.7 |
| Digital and Traditional Archives Management | 41 | 0.7 |
| Information Retrieval and Search Behavior | 40 | 0.7 |
| Innovative Human-Technology Interaction | 40 | 0.7 |
| Social and Psychological Aspects of Online Gaming | 37 | 0.6 |
| Digital Games and Media | 35 | 0.6 |

The data suggests that topics related to information literacy are the most frequent for practitioners, accounting for almost 6% of total publications. Archives are also represented, along with bibliometrics (although not as prominently as in the academic group). A notable difference from the top topics for academics is the practitioner's focus on Evidence synthesis and systematic reviews. Publications related to health sciences (*Implementation of evidence-based practice in healthcare*, *health and well-being of Arctic Indigenous peoples*, and *health sciences research and education*) comprise around 4.5% of the total.

Table 6. Top 20 topics of publications authored by practitioners

| OpenAlex topic | N. pubs | Pct. pubs |
| --- | --- | --- |
| Information Literacy in Higher Education | 83 | 2.9 |
| Library Science and Information Literacy | 82 | 2.9 |
| Usage and Impact of E-Books in Academic Settings | 54 | 1.9 |
| Implementation of Evidence-Based Practice in Healthcare | 48 | 1.7 |
| Social Inclusion in Library Services for Newcomers | 45 | 1.6 |
| Library Collection Development and Digital Resources | 42 | 1.5 |
| Health and Well-being of Arctic Indigenous Peoples | 39 | 1.4 |
| Health Sciences Research and Education | 36 | 1.3 |
| Impact of Web 2.0 on Academic Libraries | 36 | 1.3 |
| Library Science and Administration | 33 | 1.2 |
| Archival Science and Digital Preservation | 31 | 1.1 |
| Digital and Traditional Archives Management | 28 | 1.0 |
| Bibliometric Analysis and Research Evaluation | 26 | 0.9 |
| Data Sharing and Stewardship in Science | 25 | 0.9 |
| Methods for Evidence Synthesis in Research | 23 | 0.8 |
| Web and Library Services | 22 | 0.8 |
| Research Data Management Practices | 22 | 0.8 |
| Library Science and Information Systems | 19 | 0.7 |
| Meta-analysis and systematic reviews | 21 | 0.7 |
| Children's Literature and its Impact | 19 | 0.7 |

The top topics academics and practitioners tend to collaborate on are also represented in the academic-exclusive top topics. Topics relating to scientometrics and bibliometrics research and analysis account for 7.5% of all publications. The health sciences are also well-represented (7.5%). Knowledge sharing, information retrieval and literacy, and archival science are also indicated as topics frequently collectively worked on by academics and practitioners. Around 6% of publications are associated with data management and sharing. A new addition to the top 20 topics is *Wikis in Education and Collaboration*.

Table 7. Top 20 topics of publications co-authored by academics and practitioners

| OpenAlex topic | N. pubs | Pct. pubs |
|---|---|---|
| Bibliometric Analysis and Research Evaluation | 7 | 4.8 |
| Research Data Management Practices | 5 | 3.4 |
| Scientometrics and bibliometrics research | 4 | 2.7 |
| Digital Communication and Information Studies | 4 | 2.7 |
| Semantic Web and Ontologies | 4 | 2.7 |
| Health and Well-being of Arctic Indigenous Peoples | 4 | 2.7 |
| Health Sciences Research and Education | 4 | 2.7 |
| Data Sharing and Stewardship in Science | 4 | 2.7 |
| Wikis in Education and Collaboration | 4 | 2.7 |
| Knowledge Management and Sharing | 3 | 2.1 |
| Knowledge Sharing in Virtual Communities | 3 | 2.1 |
| Library Collection Development and Digital Resources | 3 | 2.1 |
| Impact of Web 2.0 on Academic Libraries | 3 | 2.1 |
| Health Policy Implementation Science | 3 | 2.1 |
| Image Retrieval and Classification Techniques | 3 | 2.1 |
| Information Literacy in Higher Education | 2 | 1.4 |
| Library Science and Information Literacy | 2 | 1.4 |
| Archival Science and Digital Preservation | 2 | 1.4 |
| Library Science and Administration | 2 | 1.4 |
| Information Retrieval Techniques and Evaluation | 2 | 1.4 |

**5.4 Citations**

Examining citation links within the dataset reveals that academics and practitioners more frequently cite works within their groups (Table 8). This is particularly true of academics.

Table 8. Number and percentage of intra-group and cross-group citations by academics and practitioners.

| Citing group | Cited group | | |
|---|---|---|---|
| | Academic | Collaboration | Practitioner |
| Academic | 1,468 (89.5%) | 18 (1.1%) | 154 (9.4%) |
| Collaboration | 35 (63.6%) | 1 (1.8%) | 19 (34.5%) |
| Practitioner | 212 (36.5%) | 23 (3.9%) | 346 (59.3%) |

One might hypothesize that the observed concentration of intra-group citations may be due to the different topics or venues in which each group is engaged. Using the venue as a proxy for the research area, we analyzed the 231 citations for which the citing venue and the cited venue were the same. This did not reduce the intra-group citation concentration but increased it, with 89.5% of citations by academics going to other academics and 59.3% of citations by practitioners going to other practitioners. This divide between academic and practitioner research is further illustrated in the giant component of the author citation network (Figure 2), though each group appear throughout the network, albeit more sparsely.

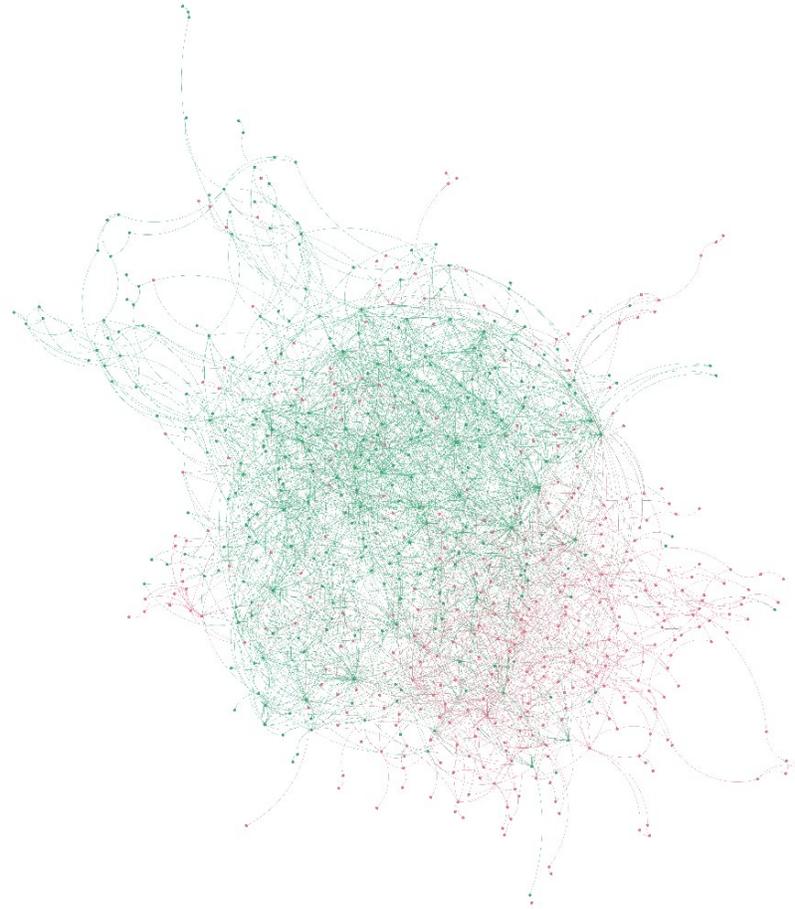

Figure 2. Citation network of Canadian academics (green nodes) and practitioners (pink nodes).

## 6. Discussion

*Types of outputs*

Our analysis of publications by academics and librarians shows that, expectedly, academics publish at higher rates than librarians. This is likely because research is a key component of their workload and vital to career advancement. Expectations of librarians to conduct research may lack definition or are not codified in workload policies or collective agreements, or librarians may not be provided the training, time, or institutional support to pursue research activities. Harrington and Gerolami (2014) highlight the high variance in collective agreements' commitment to librarians' research activities at Canadian institutions. We also find that academics publish more peer-reviewed articles, accounting for 70% of all research output. Librarians tend to favour articles, reviews, and book reviews but publish the latter two at higher rates than academics. The two groups collaborate the most on articles, reviews, and conference papers. Interestingly, academics publish conference papers more frequently, though this is an area where librarians have historically been most active (Hoffman et al., 2023; Sugimoto et al., 2014). This finding may, however, be a false correlation because academics publish at higher rates, as indicated through our results in which academics account for almost 70% of all research output, which may also result from academics' higher rate of student supervision in which they may act as co-authors of supervisee's work.

*Publication venues*

The types of output each group publishes are reflected in their choice of venue. Literature has found that librarians tend to focus on pragmatic research more relevant to practical knowledge (Hall & McBain, 2014; Nguyen & Hilder, 2018), while academics are more frequently theorists (Clayton, 1992; Genoni et al., 2006). This is reflected in our results, as librarians tend to favour venues that are focused on librarianship as a profession (e.g. References and User Services Quarterly, The Serials Librarian, The Journal of Academic Librarianship), while academics publish in a range of venues not limited to LIS. As reflected in their most frequent outputs, librarians publish in venues focused on reviews more often than academics. Academics are topically diversified, working in areas ranging from scientometrics to marine resources to computer science, as previously identified in the literature (Figuerola et al., 2017; Larivière et al., 2012; Vakkari, 2024).

*Topical overlap indicated by top topics and venues*

Librarians also tend to focus more on the health sciences than academics. This is an area where collaboration between academics and librarians has traditionally been high, typically due to the high rate of systematic reviews produced by researchers in the health sciences supported by librarians. This is shown in our results: venues in which academics and practitioners publish together include several health sciences venues (e.g., *Journal of the Medical Library Association*, *BMC Medicine*, *BMC Medical Research Methodology, BMJ Open, Journal of the Canadian Health Libraries Association*) and evidence synthesis and meta-analysis comprise approximately 1.5% of total publications, with an additional 4.5% of publications relating to health sciences topics. While librarians tend to support research teams and collaborate more in the health sciences, it is often with faculty in the health sciences; more needs to be understood about how LIS faculty collaborate with librarians in health and medicine-related fields. This may be due to the advanced research methodologies utilized in these areas, which may draw more interest or provide utility to LIS researchers.

Though Hildreth and Aytac (2007) found close topical alignment between each group, their work tends to develop independently. Notably, collaboration is observed in a few different areas. Top venues for academics include the area of archives, which is absent from practitioners' top venues. However, practitioners tend to work on archives-related topics 2% and collaborate with academics at the rate of 1.4% of publications; several archives' venues are present in the top venues where the groups collaborate. Archival science is an area where practice tends to inform theory and vice-versa. This may explain why it is a topical interface where the two groups engage, similar to information retrieval, information literacy, and library science more generally. The same applies to cataloguing and classification, documentation, and collections, which are research areas heavily informed by practice but also benefit from the advancement of theory.

Interestingly a turn towards the digital and data-related topics can be observed (Table 7). Academics and practitioners are co-working on studies related to data management and sharing in science, and digital resources, virtual communities, semantic web and web 2.0, digital preservation, and digital communication all appear within top topics, indicative of the digital transformation of Library and Information Science over the past decades.

Hearteningly, librarians and scholars tend to collaborate most frequently on topics that seem to respond to societal issues which can be addressed through the advancement of knowledge to then be applied directly in practice. This is shown by topics such as the health and well-being of Arctic Indigenous peoples, health policy implementation science, and information literacy. Further, the two groups seem to do the same for the practical manifestation of information science where knowledge is directly mobilized: the library, as indicated by a focus on topics relating to library administration, collections and resources development, information retrieval, and topics relating to education and collaboration.

*Engagement and collaboration between groups*

Table 8 shows that each group tends to cite others within the same group at high rates. Practitioners also tend to cite academics more often than the reverse, though librarians cite their group more often. This may indicate that librarians draw on work produced by academics almost as much as they derive benefits from practically focused contributions from other librarians. Practitioners also tend to cite collaborative research between the two groups more than academics do.

The network depicted in Figure 2 shows distinct academic-librarian groupings. Academics outnumber librarian researchers, though members of each group are located throughout the network, indicating that the two groups may not be cleanly siloed. Collaboration between academics and librarians is low (1.8% of all publications in this study represent collaborations between the two groups). It can be posited that each group tends to collaborate more with their respective group (academics with academics and practitioners with practitioners). The dataset used for this analysis does not possess information which allows us to observe *who* academics and practitioners collaborate with on scholarly publications not produced by collaboration within their two groups, Librarians are likely collaborating with faculty, but rather those related to the discipline in which they liaise instead of LIS faculty, supporting the objectives of their professional responsibilities and research service delivery. Collaboration between LIS faculty and librarians may not be sought by some librarians for these reasons, though past literature has established that collaboration between the two groups is highly beneficial for the advancement of LIS as a discipline and can be further promoted (Ponti, 2013; Petras, 2023).

# 7. Conclusion

This study found that librarians actively use academic work but not so much the reverse. The theory-practice divide, a longstanding discussion in LIS, has posited that academics may not be producing work that is useful or relevant to practice, furthering the gulf between these two research groups. However, it is apparent in our findings that librarians do find utility in work authored solely by academics, though academics are not drawing on work by librarians as frequently. This may be due to academics' higher volume of publications, making their work disproportionately cited. The output type could also be a factor since librarians publish reviews more frequently than academics. Academics may not inform their work with more practical perspectives due to a lack of visibility of librarians' work. We hope the dataset used in this analysis will correct this lack of exposure and facilitate research informed by practice, as well as the reverse.

This study aimed to address through its analysis whether silos can be characterized in Canadian LIS research. While there is a noticeable distinction between the two groups regarding authorship, interactions occur perhaps more frequently than previous studies have found. While the topics academics and librarians collaborate on are generally distinct from each of their foci, venues in which they publish together tend to overlap with each group's top choices. It is notable that the top topics on which academics and practitioners most frequently collaborate necessarily wed both theory and practice; this perhaps confirms a recognition between the groups that research and practice act as bi-directional pipelines, where the improvement of each can be advanced more quickly by working together and harnessing the knowledge of the other.

## 7.1 Limitations

The limitations of this research primarily stem from the database used. The database encompasses publications that have been uploaded or referenced online. However, it may not include many articles, particularly older ones, that are not listed on the web pages of librarians and educators. Also, professional conferences that do not publish proceedings do not appear in our dataset, even though the literature indicates that this is an area of dissemination favoured by practitioners (Hoffmann et al., 2023; Sugimoto et al., 2014). Additionally, the extent of the data relies on the publications mentioned in the scholars' and practitioners' curricula vitae. It is noteworthy that librarians are increasingly disseminating their research findings through unconventional channels such as blogs or websites (Finlay et al., 2013). While efforts have been made to compile these contributions, their inclusion remains uncertain as they are not typically found in traditional databases or listed in curricula vitae. Furthermore, the database primarily includes publications by librarians and professors currently in their positions at the time of writing. Retired individuals or those who have relocated to another country may not have been included in the database. Moreover, the scope of this research was limited to academic librarians, since research is included in the duties of their position. In doing so, we did not capture the professional research carried out by archivists, who do not possess the same type of status as academic librarians. Lastly, the database solely encompasses regular professors, thereby excluding the scholarly output of doctoral students, lecturers, and postdoctoral researchers.

## 7.2 Contribution

This study mapped the scholarly contributions of scholars and academic librarians in the Canadian LIS field. As an increasingly interdisciplinary field concerned with both theory and practice, this study provides a useful understanding of whether academics and practitioners collaborate on what topics and offers a sense of their awareness of the work of each group. Depicting the current state of the field opens possibilities for future exchange and collaboration and highlights research opportunities.

### 7.3 Further research

Qualitative analyses would provide a deeper understanding of the bibliometric findings presented in this study. From an information behaviour lens, understanding how librarians and practitioners collaborate and cite one another's work would offer more practical strategies for enhancing awareness of each group's respective research.

Future updates to the dataset, including the status of librarians and faculty at the time of publication, could allow for studies of their mobility as they move across institutions, as well as studies of inter-institutional collaboration across Canadian LIS schools.

The affordances of the Canadian Publications in LIS dataset would allow for a future replication of this study, against which this study could be used as a benchmark to understand how the field of LIS has evolved. It could examine how topics, academic-practitioner collaboration, publication venues, and the impact of LIS research has evolved since the publication of this study.

## 8. Conflicts of interest

There are no conflicts of interest in this project.

## 9. Acknowledgements

The authors acknowledge the support of the Social Sciences and Humanities Research Council of Canada (SSHRC), the Canadian Association of Research Libraries (CARL), the Maritime Institute for Science and Technology Studies, and the École de bibliothéconomie et des sciences de l'information (EBSI) of the Université de Montréal.

Les auteurs remercient le Conseil de recherches en sciences humaines du Canada (CRSH), l'Association des bibliothèques de recherche du Canada (ABRC), l'Institut maritime pour l'étude des sciences et technologies et l'École de bibliothéconomie et des sciences de l'information (EBSI) de l'Université de Montréal de leur soutien.

The authors wish to thank Chantal Ripp, Research Librarian (Data) at the University of Ottawa, for her helpful feedback and comments on this study.

The authors also extend their gratitude to Julia Crowell, Kellie Dalton, Brandon Fitzgibbon, Catherine Gracey, Joanna Hiemstra, Vinson Li, Kydra Mayhew, and Marc-André Simard for their contributions to building and cleaning the dataset used in this study.

# 10. Author contributions

**J-S. S.** Conceptualization, Data curation, Funding acquisition, Investigation, Methodology, Project administration, Resources, Supervision, Validation, Writing – original draft, Writing – review & editing. **M. H.** Data curation, Formal Analysis, Writing – original draft, Writing – review & editing. **G. K.** Data curation, Formal Analysis, Investigation, Methodology, Validation, Visualization, Writing – original draft, Writing – review & editing. **P. R.** Data curation, Visualization, Writing – original draft, Writing – review & editing. **C. P.** Data curation, Writing – original draft, Writing – review & editing. **P. M.** Conceptualization, Data curation, Formal Analysis, Funding acquisition, Investigation, Methodology, Project administration, Resources, Supervision, Validation, Visualization, Writing – original draft, Writing – review & editing.

# 11. Data availability

The data set and R scripts used to produce it are available on Zenodo: 10.5281/zenodo.11355851.